\documentclass[preprint,aps, nofootinbib]{revtex4}
\input{epsf.tex}
\usepackage[dvips]{graphics,epsfig}
\def\be{\begin{equation}}
\def\ee{\end{equation}}
\def\ba{\begin{eqnarray}}
\def\ea{\end{eqnarray}}

\def\dalemb#1#2{{\vbox{\hrule height.#2pt
        \hbox{\vrule width.#2pt height#1pt \kern#1pt \vrule width.#2pt}
        \hrule height.#2pt}}}

\def\dalemb#1#2{{\vbox{\hrule height.#2pt
        \hbox{\vrule width.#2pt height#1pt \kern#1pt \vrule width.#2pt}
        \hrule height.#2pt}}}

\def\ba{\begin{eqnarray}}
\def\ea{\end{eqnarray}}
\def\be{\begin{equation}}
\def\ee{\end{equation}}

\def\gtorder{\mathrel{\raise.3ex\hbox{$>$}\mkern-14mu
             \lower0.6ex\hbox{$\sim$}}}
\def\ltorder{\mathrel{\raise.3ex\hbox{$<$}\mkern-14mu
             \lower0.6ex\hbox{$\sim$}}}

\def\to{\rightarrow}

\def\PL{{\it Phys. Lett.} }

\def\be{\beta}

\def\frac#1#2{{\textstyle{{#1}\over {#2}}}}

\def\lsim{\mathrel{\rlap{\lower4pt\hbox{\hskip1pt$\sim$}}
    \raise1pt\hbox{$<$}}}
\def\gsim{\mathrel{\rlap{\lower4pt\hbox{\hskip1pt$\sim$}}
    \raise1pt\hbox{$>$}}}
\def\sqr#1#2{{\vcenter{\vbox{\hrule height.#2pt
         \hbox{\vrule width.#2pt height#1pt \kern#1pt
         \vrule width.#2pt}
         \hrule height.#2pt}}}}

\begin{document}

\rightline{DF/IST-4.2006}

\title{
Lorentz Symmetry Derived from Lorentz Violation in the Bulk
}

\author{Orfeu Bertolami}
\altaffiliation{Email address: orfeu@cosmos.ist.utl.pt}

\author{Carla Carvalho}
\altaffiliation{Email address: ccarvalho@ist.edu}

\vskip 0.5cm

\affiliation{Departamento de F\'\i sica, Instituto Superior T\'ecnico \\
Avenida Rovisco Pais 1, 1049-001 Lisboa, Portugal}

\vskip 1.0cm

\begin{abstract}%
{
We consider bulk fields coupled to the graviton in a Lorentz violating
fashion. We expect that the overly tested Lorentz symmetry might set
constraints on the induced Lorentz violation in the brane, and hence
on the dynamics of the interaction of bulk fields on the brane.
We also use the requirement for Lorentz symmetry to constrain the 
cosmological constant observed on the brane
}
\end{abstract}

\maketitle

\section{Introduction}

Lorentz invariance is one of the most well tested symmetries of physics. 
Nevertheless, the possibility of violation of this invariance 
has been widely discussed in the recent literature (see 
e.g. \cite{Kostelecky1}). Indeed, the spontaneous breaking of Lorentz
symmetry may arise in the context of string/M-theory due to the existence of
non-trivial solutions in string field theory  \cite{Kostelecky2}, 
in loop quantum gravity \cite{Gambini}, 
in noncommutative field theories 
\cite{Carroll,Bertolami} or via the spacetime variation of 
fundamental couplings \cite{Lehnert}. 
This putative breaking has also implications in ultra-high energy cosmic 
ray physics \cite{Sato,Bertolami1}. Lorentz violation modifications
of the dispersion  
relations via five dimensional operators for fermions have also been
considered  
and constrained \cite{Bertolami2}.  It has also been speculated that
Lorentz symmetry is connected with  
the cosmological constant problem \cite{Bertolami3}. However, the main
conclusion of these  
studies is that Lorentz symmetry holds up to about
$2\times 10^{-25}$ \cite{Kostelecky1,Bertolami1}.

Efforts to examine a putative breaking of Lorentz invariance
have been concerned mainly with the phenomenological aspects of the 
spontaneous breaking of Lorentz symmetry in particle physics and only
recently have the implications for gravity been more
closely studied \cite{Kostelecky3,BParamos}. The idea is to consider a 
vector field coupled to gravity that undergoes 
spontaneous Lorentz symmetry breaking by acquiring a vacuum 
expectation value in a potential.

Moreover, recent developments in string theory suggest that we may 
live in a braneworld embedded in a higher dimensional universe. In the 
context of the Randall-Sundrum cosmological models, the warped geometry of the
bulk along the extra spacial dimension suggests an anisotropy
which could be associated with the breaking of the bulk Lorentz symmetry.

In this paper we study how spontaneous Lorentz violation in the bulk
repercusses on the brane and how it can be constrained.
We consider a vector field in the bulk 
which acquires a non-vanishing expectation
value in the vacuum and introduces spacetime anisotropies in the
gravitational field equations through the coupling with the graviton.
For this purpose, we derive the field equations and project them parallel and
orthogonally to the brane. We then establish how to derive brane quantities
from bulk quantities by adopting Fermi normal coordinates with respect
to the directions on the brane and continuing into the bulk using the
Gauss normal prescription.

We parameterize the worldsheet in terms of coordinates
$x^{A}=(t_b,{\bf x}_b)$ 
intrinsic to the brane. Using the chain rule, we may express the brane
tangent and normal unit vectors in terms of the bulk basis as follows:
\begin{eqnarray}
\hat e_A &=&{\partial \over \partial x^A}
=X_A^\mu{\partial \over \partial x^\mu} =X_A^\mu \hat e_\mu,\cr
\hat e_N &=&{\partial \over \partial n}
=N^\mu {\partial \over \partial x^\mu} =N^\mu \hat e_\mu,
\end{eqnarray}
with
\begin{eqnarray}
g_{\mu\nu}N^\mu N^\nu =1,\quad g_{\mu\nu}N^\mu X^\nu_A =0,
\end{eqnarray} 
where ${\bf g}$ is the bulk metric
\begin{eqnarray}
{\bf g}
&=&g_{\mu\nu}{\hat e}_\mu\otimes {\hat e}_\nu 
=g_{AB}~{\hat e}_A\otimes {\hat e}_B
+g_{AN}~{\hat e}_A\otimes {\hat e}_N 
+g_{NB}~{\hat e}_N\otimes {\hat e}_B
+g_{NN}~{\hat e}_N\otimes {\hat e}_N
\end{eqnarray}
To obtain the metric induced on the brane we expand the bulk basis
vectors in terms of the coordinates intrinsic to the brane and keep the
doubly brane tangent components only. It follows that 
\begin{eqnarray}
g_{AB} =X^\mu _AX^\nu _B~g_{\mu\nu}
\end{eqnarray}
is the $(3+1)$-dimensional
metric induced on the brane by the $(4+1)$-dimensional bulk metric
$g_{\mu\nu}.$ 
The induced metric with upper indices is defined by the relation
\begin{eqnarray}
g_{AB}~g^{BC}=\delta _A{}^C.
\end{eqnarray}
It follows that we can write any bulk tensor field as a linear
combination of mutually orthogonal vectors on the brane, $\hat e_A,$ 
and a vector normal to the brane, $\hat e_N.$ We illustrate the
example of a vector $B_{\mu}$ and a tensor $T_{\mu\nu}$ bulk fields as follows
\begin{eqnarray}
{\bf B}
&=& B_{A}~{\hat e}_A +B_{N}~{\hat e}_N, \\
{\bf T} 
&=&T_{AB}~{\hat e}_A\otimes {\hat e}_B
+T_{AN}~{\hat e}_A\otimes {\hat e}_N 
+T_{NB}~{\hat e}_N\otimes {\hat e}_B
+T_{NN}~{\hat e}_N\otimes {\hat e}_N.
\end{eqnarray}
Derivative operators decompose similarly.
We write the derivative operator $\nabla$ as
\begin{eqnarray}
\nabla =(X^\mu_A +N^\mu)\nabla_\mu =\nabla_A +\nabla_N.
\end{eqnarray}
Fixing a point on the boundary, we introduce coordinates for the neighborhood
choosing them to be Fermi normal.
All the Christoffel symbols of the metric on the boundary are
thus set to zero, although the partial derivatives do not in general
vanish. The non-vanishing connection coefficients are
\begin{eqnarray}
\nabla_A {\hat e}_B &=&-K_{AB}~{\hat e}_N, \cr
\nabla_A {\hat e}_N &=&+K_{AB}~{\hat e}_B, \cr
\nabla_N {\hat e}_A &=&+K_{AB}~{\hat e}_B, \cr
\nabla_N {\hat e}_N &=&0,
\end{eqnarray}
as determined by the Gaussian normal prescription for the continuation
of the coordinates off the boundary.
For the derivative operator $\nabla\nabla$ we find that
\begin{eqnarray}
\nabla\nabla&=& g^{\mu\nu}\nabla_\mu \nabla_\nu \cr
&=&g^{AB}\left[ (X^\mu_A\nabla_\mu)(X^\nu_B\nabla_\nu) 
-X^\mu_A(\nabla_\mu X^\nu_B)\nabla_\nu \right] 
%&+&g^{AN}\left[ (X^\mu_A\nabla_\mu)(N^\nu\nabla_\nu) 
%-(X^\mu_A\nabla_\mu)(N^\nu\nabla_\nu) \right] \cr
%&+&g^{NB}\left[ (N^\mu\nabla_\mu)(X^\nu_B\nabla_\nu) 
%-(N^\mu\nabla_\mu)(X^\nu_B\nabla_\nu) \right] \cr
+g^{NN}\left[ (N^\mu\nabla_\mu)(N^\nu\nabla_\nu) 
-N^\mu(\nabla_\mu N^\nu)\nabla_\nu \right] \cr
&=&g^{AB}\left[ \nabla_A \nabla_B +K_{AB}\nabla_N\right]
+               \nabla_N \nabla_N.
\end{eqnarray}
We can now decompose the Riemann tensor, $R_{\mu\nu\rho\sigma},$ along
the tangent and the normal directions to the surface of the brane as
follows
\begin{eqnarray}
R_{ABCD} &=& R_{ABCD}^{(ind)} +K_{AD}K_{BD} -K_{AC}K_{BD}, \qquad\\
R_{NBCD} &=& K_{BC;D} -K_{BD;C}, \\
R_{NBND} &=& K_{BC}K_{DC} -K_{BC,N},
\end{eqnarray}
from which we find the decomposition of the Einstein tensor,
$G_{\mu\nu},$ obtaining the Gauss-Codacci relations
\begin{eqnarray}
G_{AB} &=&G_{AB}^{(ind)} +2K_{AC}K_{CB} -K_{AB}K -K_{AB,N} 
-{1\over 2}g_{AB}\left( 3K_{CD}K_{DC} -K^2 -2K_{,N}\right),\quad \\
G_{AN} &=&K_{AB;B} -K_{;A}, \\
G_{NN} &=&{1\over 2}\left( -R^{(ind)} -K_{CD}K_{DC} +K^2\right).
\label{eqn:gc}
\end{eqnarray}

\section{Bulk Vector Field Coupled to Gravity}

We consider a bulk vector field ${\bf B}$ with a non-trivial coupling 
to the graviton in a five-dimensional anti-de Sitter space. 
The Lagrangian density consists of
the Hilbert term, the cosmological constant term,
the kinetic and potential terms for 
${\bf B}$ and the ${\bf B}$--graviton interaction term, as follows
\begin{eqnarray}
{\cal L} 
={1\over {\kappa_{(5)}^2}}R -2\Lambda
+\lambda B^{\mu}B^{\nu}R_{\mu\nu}
-{1\over 4}B_{\mu\nu}B^{\mu\nu} -V(B^\mu B_\mu \pm b^2 ),\quad
\end{eqnarray}
where $B_{\mu\nu} =\nabla_{\mu}B_{\nu} -\nabla_{\nu}B_{\mu}$ is the
tensor field associated with $B_{\mu}$ and $V$ is the potencial which 
induces the breaking of Lorentz symmetry once the ${\bf B}$ field is
driven to  the minimum at $B^\mu B_\mu \pm b^2 = 0 $, $b^2$ being  
a real positive constant. As discussed in the introduction, 
this model has been proposed in order to analyse 
the impact on the gravitational 
sector of the breaking of Lorentz symmetry \cite{Kostelecky3,BParamos}.
Furthermore, in the present model 
$\kappa_{(5)}^2=8\pi G_{N}=M_{Pl}^3,$
$M_{Pl}$ is the five-dimensional Planck mass
and $\lambda$ is a dimensionless coupling constant that we have
inserted to track the effect of the interaction.
In the cosmological constant term
$\Lambda =\Lambda_{(5)} +\Lambda_{(4)}$ we have included both the bulk
vacuum value $\Lambda_{(5)}$ and that of the brane $\Lambda_{(4)},$
described by a brane tension $\sigma$ localized on the locus of the
brane, $\Lambda_{(4)}=\sigma\delta(N).$

By varying the action with respect to the metric, we 
obtain the Einstein equation
\begin{eqnarray}
{1\over \kappa_{(5)}^2}G_{\mu\nu} +\Lambda g_{\mu\nu}
-\lambda L_{\mu\nu} -\lambda \Sigma_{\mu\nu} ={1\over 2}T_{\mu\nu},
\end{eqnarray}
where  
\begin{eqnarray}
L_{\mu\nu} &=&{1\over 2}g_{\mu\nu}B^{\rho}B^{\sigma}R_{\rho\sigma} 
-\left( B_{\mu}B^{\rho}R_{\rho\nu} +R_{\mu\rho}B^{\rho}B_{\nu}\right),\qquad\\
\Sigma_{\mu\nu} &=& {1\over 2}\bigl[
\nabla_{\mu}\nabla_{\rho}(B_{\nu}B^{\rho}) 
+\nabla_{\nu}\nabla_{\rho}(B_{\mu}B^{\rho})\
-\nabla^2(B_{\mu}B_{\nu}) 
-g_{\mu\nu}\nabla_{\rho}\nabla_{\sigma}(B^{\rho}B^{\sigma})\bigr]
\end{eqnarray}
are the contributions from the interaction term and
\begin{eqnarray}
T_{\mu\nu} = 
B_{\mu\rho}B_{\nu}{}^{\rho} +4V^\prime B_{\mu}B_{\nu}
+g_{\mu\nu}\left[ -{1\over 4}B_{\rho\sigma}B^{\rho\sigma} -V\right] \quad
\end{eqnarray}
is the contribution from the vector field for the stress-energy tensor.
For the equation of motion for the vector field ${\bf B},$ obtained by
varying the action with respect to $B_{\mu},$ we find that
\begin{eqnarray}
\nabla^{\nu}
 \left( \nabla_{\nu}B_{\mu} -\nabla_{\mu}B_{\nu}\right)
-2V^{\prime}B_{\mu}
+2\lambda B^{\nu}R_{\mu\nu} 
=0, \quad
\end{eqnarray}
where $V^{\prime} =dV/dB^2$.

We now proceed to project the equations parallelly and orthogonally to
the surface of the brane. 
Following the prescription used in the derivation of the Gauss-Codacci
relations, we derive the components of the stress-energy tensor and of
the interaction terms.  Similarly,
the equation of motion for the vector field ${\bf B}$ decomposes as follows 
\begin{eqnarray}
&&\nabla_{C}\left( \nabla_{C}B_{A} -\nabla_{A}B_{C}\right) 
+\nabla_{N}\left( \nabla_{N}B_{A} -\nabla_{A}B_{N}\right)\cr
&+&2K_{AC}\left( \nabla_{C}B_{N} -\nabla_{N}B_{C}\right)
+K\left( \nabla_{N}B_{A} -\nabla_{A}B_{N}\right)
-2V^{\prime}B_{A}\cr 
&+&2\lambda \Bigl[ 
B_{C}\left( R^{(ind)}_{AC} +2K_{AD}K_{DC} -K_{AC}K -K_{AC,N}\right)
+B_{N}\left( K_{AC;C} -K_{;A}\right) \Bigr]
=0,
\label{eqn:B_A}
\\
\cr
&&\nabla_{C}\left( \nabla_{C}B_{N} -\nabla_{N}B_{C}\right) 
-2V^{\prime}B_{N}\cr 
&+&2\lambda \left[ 
B_{C}\left( K_{CD;D} -K_{;C}\right)
+B_{N}\left( K_{CD}K_{CD} -K_{;N}\right) \right] = 0,
\label{eqn:B_N}
\end{eqnarray}
parallelly and orthogonally to the brane respectively, which we
include here for the purpose of illustration.

Next we proceed to derive the induced equations of motion for both
the metric and the vector field in terms of quantities measured on the
brane. The induced equations on the brane are the (AB) projected
components after the singular terms across the brane are subtracted
out by the substitution of the matching conditions.
Considering the brane as a $Z_2$-symmetric shell of
thickness $2\delta$ in the limit $\delta \to 0,$ 
derivatives of quantities discontinuous across the brane generate
singular distributions on the brane. Integration of these terms in the
coordinate normal to the brane
relates the induced geometry %on the brane 
with the localization of the induced stress-energy %on the brane 
in the form of matching conditions.
First we consider the Einstein equations and then the equations of
motion for ${\bf B}$ which, due to the coupling of ${\bf B}$ to
gravity, will also be used as complementary conditions for the dynamics of
the metric on the brane.

Combining the Gauss-Codacci relations with the projections of the
stress-energy tensor and the interaction source terms, we integrate
the $(AB)$ component of the Einstein equation in the 
coordinate normal to the brane to obtain the matching conditions for
the extrinsic curvature across the brane, i.e. the Israel matching conditions.
From the $Z_{2}$ symmetry it follows that 
$B_{A}(-\delta) =+B_{A}(+\delta)$ 
but that $B_{N}(-\delta) =-B_{N}(+\delta),$ and consequently that 
$(\nabla_{N}B_{A})(-\delta) =-(\nabla_{N}B_{A})(+\delta)$  and 
$(\nabla_{N}B_{N})(-\delta) =+(\nabla_{N}B_{N})(+\delta)$.
Moreover,
$g_{AB}(N=-\delta) =+g_{AB}(N=+\delta)$ implies that 
$K_{AB}(N=-\delta) =-K_{AB}(N=+\delta)$.
Hence, we find for the $(AB)$ matching conditions that
\begin{eqnarray}
&&{1\over \kappa _{(5)}^2}
\left[ -K_{AB} +g_{AB}K\right]%^{+\delta}_{-\delta}
\cr
&=&{1\over 2}\int ^{+\delta}_{-\delta} dN\left[ 
-g_{AB}\Lambda_{(4)}\right]%^{+\delta}_{-\delta}
\cr
&+&{\lambda \over 2}\Big[
\nabla_{A}(B_{B}B_{N}) +\nabla_{B}(B_{A}B_{N})
-\nabla_{N}(B_{A}B_{B}) \cr
&&{}+4( B_{A}B_{C}K_{CB} +K_{AC}B_{C}B_{B})
-2K_{AB}B_{N}B_{N}\cr
&&{}+g_{AB}\left( 
-2\nabla_{C}(B_{C}B_{N}) -\nabla_{N}(B_{N}B_{N}) 
+K_{CD}B_{C}B_{D} -KB_{N}B_{N}\right)
\Big].%^{+\delta}_{-\delta}.
\label{eqn:imc}
\end{eqnarray}
These provide boundary conditions for ten of the fifteen degrees of
freedom. Five additional boundary conditions are provided by the
junction conditions for the $(AN)$ and 
$(NN)$ components of the projection of the Einstein equations.
From inspection of the $(AN)$ component, we note that 
\begin{eqnarray}
G_{AN} &=& K_{AB;B} -K_{;A}\cr 
&=& -\nabla_{B}\left( \int ^{+\delta}_{-\delta}
dN~G_{AB}\right)
=-\kappa_{(5)}^2 \nabla_{B}{\cal T}_{AB} =0
\end{eqnarray}
which vanishes due to conservation of the induced stress-energy tensor
${\cal T}_{AB}$ on the brane.
From integration of the $(NN)$ component in the normal direction to the brane,
we find the following junction condition
\begin{eqnarray}
\nabla_{C}(B_{C}B_{N}) +3KB_{N}B_{N} -K_{CD}B_{C}B_{D} =\sigma, 
\label{eqn:G_NN:junction}
\end{eqnarray}
which we substitute back in, obtaining
\begin{eqnarray}
&&{1\over \kappa_{(5)}^2}{1\over 2}
\left( -R^{(ind)} -K_{CD}K_{CD} +K^2\right)\cr
&=&{1\over 2}
\left[ -{1\over 4}B_{CD}B_{CD}  -V\right] 
\cr
&+&{1\over 2}\Bigl[
-\nabla_{C}\nabla_{D}(B_{C}B_{D})
-\nabla_{C}\nabla_{C}(B_{N}B_{N})
+12B_{N}\nabla_{C}\nabla_{C}B_{N} -20V^{\prime}B_{N}B_{N}\cr
&&{}+2\left( K_{CD} -g_{CD}K\right)\nabla_{C}(B_{D}B_{N})
+2K_{CD}B_{D}(\nabla_{C}B_{N})%\cr
%&&{}
+K_{CD;C}B_{D}B_{N}\cr
&&{}
+\left( 7K_{CD}K_{CD} -K^2\right)B_{N}B_{N}
+\left( 7K_{CE}K_{ED} +KK_{CD}\right)B_{C}B_{D}\Bigr].
\label{eqn:G_NN:induced}
\end{eqnarray}
However, the Israel matching conditions also contain terms which depend on the
prescription for the continuation of ${\bf B}$ 
out of the brane and into the bulk, 
namely $\nabla_{N}B_{A}$ and $\nabla_{N}B_{N}$. The five
additional boundary conditions required are those for the vector field
{\bf B.}
In Ref.~\cite{BuCarvalho05} 
the boundary conditions for bulk fields were derived subject to 
the condition that 
modes emitted by the brane into the bulk  
do not violate the gauge defined in the bulk.
Here, however,  
we integrate the (A) and (N) components of the equation of
motion for ${\bf B},$ Eq.~(\ref{eqn:B_A}) and Eq.~(\ref{eqn:B_N})
respectively, to find the corresponding junction condition for $B_{A}$
and for $B_{N}$ across the brane. 
From Eq.~(\ref{eqn:B_A}) we have that
\begin{eqnarray}
\int ^{+\delta}_{-\delta} dN \bigl[ 
\nabla_{N}\left(
\nabla_{N}B_{A} -\nabla_{A}B_{N}\right)
-2K_{AC}(\nabla_{N}B_{C}) 
%-2V_{(4)}^{\prime}B_{A}
-2\lambda B_{C}K_{AC,N}\bigr] 
=0.
\end{eqnarray}
If $\delta$ is sufficiently small, the difference between $K_{AB;N}$
and $K_{AB,N}$ is negligibly small. It follows that, in the limit where
$\delta \to 0,$ we can assume that $\nabla_{N}\approx \partial_{N}$. It 
then follows that
\begin{eqnarray}
\nabla_{N}B_{A} -\nabla_{A}B_{N} 
-2K_{AC}B_{C}
%-\int ^{+\delta}_{-\delta}dN~( V_{(4)}^{\prime}B_{A})
=0.
\label{eqn:B_A:junction}
\end{eqnarray}
Similarly, from Eq.~(\ref{eqn:B_N}) we find that
\begin{eqnarray}
\int ^{+\delta}_{-\delta} dN \left[ 
-\nabla_{N}\nabla_{C}B_{C} 
%-2V_{(4)}^{\prime} B_{N} 
-\lambda \nabla_{N}(K B_{N})\right]=0,
%&\Leftrightarrow&
\end{eqnarray}
which becomes
\begin{eqnarray}
\nabla_{C}B_{C} +\lambda K B_{N} 
%+\int ^{+\delta}_{-\delta} dN~( V_{(4)}^{\prime}B_{N})
=0.
\label{eqn:B_N:junction}
\end{eqnarray}
The junction conditions Eq.~(\ref{eqn:B_A:junction}) and 
Eq.~(\ref{eqn:B_N:junction}) offer the required $(4+1)$ boundary
conditions respectively for $B_{A}$ and $B_{N}$ on the brane. 
Substituting the junction condition for $B_{A}$
back in Eq.~(\ref{eqn:B_A}) and using the result from $G_{AN}=0,$
we find for the induced equation of motion for $B_{A}$ on the brane that
\begin{eqnarray}
\nabla_{C}\left( \nabla_{C}B_{A} -\nabla_{A}B_{C}\right) 
+2K_{AC}(\nabla_{C}B_{N})
%+K\int dN~( V_{(4)}^{\prime}B_{A})
-2V^{\prime}B_{A} 
%+2(K_{AC;C} -\lambda K_{;A})B_{N}
+2\lambda 
B_{C}\left( R^{(ind)}_{AC} +2K_{AD}K_{DC}\right)
=0.
\label{eqn:B_A:induced}
\end{eqnarray}
Similarly, substituting the junction condition for $B_{N}$ 
back in Eq.~(\ref{eqn:B_N}) we obtain
\begin{eqnarray}
\nabla_{C}\nabla_{C}B_{N} 
-2V^{\prime}B_{N} 
+\lambda \left[
K(\nabla_{N}B_{N})
%+B_{C}(K_{CD;D} -K_{;C})
+B_{N}K_{CD}K_{CD} \right]
=0.
\label{eqn:B_N:induced}
\end{eqnarray}
Thus, Eq.~(\ref{eqn:B_A:junction}) provides the value at the boundary
for $\nabla_{N}B_{A}$ and Eq.~(\ref{eqn:B_N:induced}) provides that
for $\nabla_{N}B_{N}$.
Using the results derived above in the Israel matching conditions we find that
\begin{eqnarray}
&&{1\over \kappa _{(5)}^2}
\left[ -K_{AB} +g_{AB}K\right]%^{+\delta}_{-\delta}
\cr
&=&%{1\over 2}\int dN\left[ 
%{1\over 2}\left( 4V^{\prime}_{(4)}B_{A}B_{B} -g_{AB}V_{(4)}\right)
%-g_{AB}\Lambda_{(4)}\right] 
-g_{AB}~\sigma 
+{1\over 2}\Big[
(\nabla_{A}B_{B})B_{N} +(\nabla_{B}B_{A})B_{N}
%-B_{B}\int dN~(V^{\prime}_{(4)}B_{A})
%-B_{A}\int dN~(V^{\prime}_{(4)}B_{B})
\Big]\cr
&+& B_{A}B_{C}K_{CB} +K_{AC}B_{C}B_{B} -K_{AB}B_{N}B_{N}\cr
&+&g_{AB}\biggl[ 
-\nabla_{C}(B_{C}B_{N}) 
+{1\over 2}K_{CD}B_{C}B_{D} -{1\over 2}KB_{N}B_{N}\cr
&&{}+{1\over K} \bigl(
B_{N}\nabla_{C}\nabla_{C}B_{N} 
%+V^{\prime}_{(5)}B_{C}B_{C} 
-2V^{\prime}B_{N}B_{N}
%\cr
%&&{}-{1\over 2}B_{C}\nabla_{D}(\nabla_{D}B_{C} -\nabla_{C}B_{D})
%-B_{C}K_{CD}(\nabla_{D}B_{N})
%-{1\over 2}KB_{C}\int dN~(V^{\prime}_{(4)}B_{C})\cr
%&&{}-B_{C}B_{D}\left( R^{(ind)}_{CD} +2K_{CE}K_{ED}\right)
+B_{N}B_{N}K_{CD}K_{CD}\bigr)
\biggr].%^{+\delta}_{-\delta}.
\end{eqnarray}
The Israel matching conditions provide an equation for the trace of
the extrinsic curvature, $K.$
Finally, using Eq.~(\ref{eqn:G_NN:induced})
in the $(AB)$ Einstein equation, we find for the 
Einstein equation induced on the brane that
\begin{eqnarray}
&&{1\over \kappa_{(5)} ^2}\left[ 
G_{AB}^{(ind)} +2K_{AC}K_{BC} -K_{AB}K 
+{1\over 2}g_{AB}\left( -R^{(ind)} -4K_{CD}K_{CD} +2K^2\right)\right] 
+g_{AB}\Lambda_{(5)}\cr
&+&{}{1\over 2}\left[ 
-B_{AC}B_{BC} -4V^{\prime}B_{A}B_{B} 
+{1\over 2}g_{AB}\left( B_{CD}B_{CD} +6V\right)\right]\cr
%%%%%
&=&{1\over 2}g_{AB}\biggl[
%B_{C}B_{D}R^{(ind)}_{CD} \cr
-2\nabla_{C}\nabla_{D}(B_{C}B_{D})
-\nabla_{C}\nabla_{C}(B_{N}B_{N})
+12B_{N}\nabla_{C}\nabla_{C}B_{N}
-20 V^{\prime}B_{N}B_{N} 
\cr
&&{}%\quad
+4(K_{CD} -g_{CD}K)\nabla_{D}(B_{C}B_{N})
+6K_{CD}B_{D}(\nabla_{C}B_{N}) 
+K B_{C}(\nabla_{C}B_{N})\cr
&&{}%\quad
+B_{C}B_{D}R^{(ind)}_{CD}
+9K_{CD}K_{CD}B_{N}B_{N} 
+14K_{CE}K_{DE}B_{C}B_{D} %+K B_{C}\nabla_{C}B_{N}
-K\sigma
\biggr]\cr
%%%%
&+&{1\over 2}\biggl[
%-2B_{A}B_{C}\left( R^{(ind)}_{CB} +2K_{CD}K_{BD}\right)
%-2B_{B}B_{C}\left( R^{(ind)}_{CA} +2K_{AD}K_{CD}\right) \cr
%&&{}+(K_{AC;B} +K_{BC;A} -2K_{AB;C})B_{N}B_{N} \cr
\nabla_{A}\nabla_{C}(B_{B}B_{C})
+\nabla_{B}\nabla_{C}(B_{A}B_{C})
-\nabla_{C}\nabla_{C}(B_{A}B_{B})\cr
&&{}%\quad
-2K_{AC}\nabla_{C}( B_{B}B_{N}) 
-2K_{AC}(B_{B}\nabla_{C}B_{N} +B_{C}\nabla_{B}B_{N})\cr
&&{}
-2K_{BC}\nabla_{C}( B_{A}B_{N}) 
-2K_{BC}(B_{A}\nabla_{C}B_{N} +B_{C}\nabla_{A}B_{N})\cr
&&{}%\quad
+KB_{N}(\nabla_{A}B_{B} +\nabla_{B}B_{A}) 
-2K_{AB}B_{N}(\nabla_{C}B_{C}) %+3KB_{N}B_{N} -\sigma) 
\cr
&&{}-{8\over K}K_{AB}( \nabla_{C}\nabla_{C}B_{N} -2V^{\prime}B_{N} 
+B_{N}K_{CD}K_{CD})\cr
%%%%
&&{}%\quad
-2B_{A}B_{C}\left( R^{(ind)}_{CB} +2K_{CD}K_{BD}\right)
-2B_{B}B_{C}\left( R^{(ind)}_{CA} +2K_{AD}K_{CD}\right)\cr
&&{}+(K_{AC;B} +K_{BC;A} -2K_{AB;C})B_{N}B_{N} %\cr
%&&{}%\quad
+( K_{AC}B_{B} +K_{BC}B_{A})( -5K_{DC}B_{D} +KB_{C}) \cr
&&{}-6K_{AC}K_{BD}B_{C}B_{D} %\cr
-2K_{AB}( 3KB_{N}B_{N} -\sigma)
%&&{}-{8\over K}K_{AB}( \nabla_{C}\nabla_{C}B_{N} -2V^{\prime}B_{N} 
%+B_{N}K_{CD}K_{CD})
\biggr].
\end{eqnarray}
The results obtained above show both the coupling of the bulk to the
brane and the coupling of the vector field ${\bf B}$ to the geometry
of the spacetime. The first is manifested in the dependence on normal
components in the induced equations; the latter is manifested in the
presence of terms of the form $(R_{AB}B_{C}B_{D}).$ Terms of the form
$(K_{AB}B_{N})$ illustrate both couplings, where $B_{N}$ relates with
$K$ and $B_{A}$ by Eq.~(\ref{eqn:B_N:junction}). The directional
dependence on the $N$ direction is encapsulated in the extrinsic
curvature. 
In the fourth line we can substitute the Israel matching condition
found above. 
However, 
the derivatives of the extrinsic curvature
along directions parallel to the brane which appear in the twelfth line
are not reducible to quantities intrinsic to the brane.

\section{Bulk Vector Field with a Non-vanishing
Vacuum Expectation Value}

In this section we particularize the formalism developed above for
the case when the bulk vector field ${\bf B}$ acquires a non-vanishing
vacuum value by spontaneous symmetry breaking akin to the Higgs mechanism.
The vacuum value generates the breaking of the Lorentz
symmetry by selecting the direction orthogonal to the plane of the
brane.
The tangential part of the vector field with respect to
the brane will acquire an expectation value, $\left<B_{A}\right>\not=0$, 
whereas the expectation
value of the normal component is chosen, for simplicity, 
to vanish on the brane, $\left<B_{N}\right> =0$, 
as we are interested in the effect that Lorentz symmetry breaking in the bulk 
has on the brane. Choosing instead $\left<B_{A}\right> =0$ and
$\left<B_{N}\right> \not=0$ would also violate Lorentz symmetry on the
  brane. However, the implied matching conditions would be
  incompatible with the condition for the convariant conservation of
  the vacuum expectation value of the field ${\bf B},$ 
$\nabla_{A}\left<B_{C}\right> = 0$ \cite{Kostelecky3,BParamos}.
Moreover, the vacuum value is at a zero of both the potential $V$ and
its derivative $V^{\prime}$. 
The junction conditions from the equations for $B_{A},$ $B_{N},$
$G_{NN}$ and $G_{AB}$ reduce respectively to 
\begin{eqnarray}
\nabla_{N}\left<B_{A}\right> -2K_{AC}\left<B_{C}\right> &=&0\\
\nabla_{C}\left<B_{C}\right>&=&0\\
-K_{CD}\left<B_{C}\right>\left<B_{D}\right> &=&\sigma \\
{1\over \kappa_{(5)}^2}\left[ -K_{AB} +g_{AB}K\right]
&=&-g_{AB}~\sigma
+\left<B_{A}\right>\left<B_{C}\right>K_{CB}
+\left<B_{B}\right>\left<B_{C}\right>K_{AC}
\end{eqnarray}
and the induced equations of motion become
\begin{eqnarray}
\nabla_{C}\left( \nabla_{C}\left<B_{A}\right> 
-\nabla_{A}\left<B_{C}\right>\right)
+2\left<B_{C}\right>\left( R_{AC}^{(ind)} +2K_{AD}K_{DC}\right)=0
\end{eqnarray}
for $B_{A},$
\begin{eqnarray}
&&{1\over \kappa_{(5)}^2}{1\over 2}\left( 
-R^{(ind)} -K_{CD}K_{CD} +K^2\right)\cr
&=&{1\over 2}\left[
-{1\over 4}\left< B_{CD}\right>\left< B_{CD}\right>
-\nabla_{C}\nabla_{D}(\left< B_{C}\right>\left< B_{D}\right>)
+7K_{CE}K_{ED}\left< B_{C}\right>\left< B_{D}\right> -K\sigma
\right]
\end{eqnarray}
for $G_{NN}$ and finally
\begin{eqnarray}
&&{1\over \kappa_{(5)} ^2}\left[ 
G_{AB}^{(ind)} +2K_{AC}K_{BC} -{1\over 2}K_{AB}K 
+{1\over 2}g_{AB}\left( R^{(ind)} -K_{CD}K_{CD} -K^2\right)\right] 
+g_{AB}\Lambda_{(5)}\cr
&-&{1\over 2}
\left< B_{AC}\right>\left< B_{BC}\right> \cr
&=&{1\over 2}\biggl[
{1\over 4}\left< B_{A}\right>\nabla_{C}
\left( 5\nabla_{C}\left< B_{B}\right> 
-9\nabla_{B}\left< B_{C}\right>\right)
+{1\over 4}\left< B_{B}\right>\nabla_{C}
\left( 5\nabla_{C}\left< B_{A}\right>
-9\nabla_{A}\left< B_{C}\right>\right)\cr
&&{}+\nabla_{A}\nabla_{C}(\left< B_{B}\right>\left< B_{C}\right>)
+\nabla_{B}\nabla_{C}(\left< B_{A}\right>\left< B_{C}\right>)
-2(\nabla_{C}\left< B_{A}\right>)(\nabla_{C}\left< B_{B}\right>)\cr
&&{}+{5\over 2}\left<B_{A}\right>\left<B_{C}\right>R^{(ind)}_{CB}
+{5\over 2}\left<B_{B}\right>\left<B_{C}\right>R^{(ind)}_{AC}
-6K_{AC}K_{BD}\left< B_{C}\right>\left< B_{D}\right>
+2K_{AB}\sigma \biggr]\cr
&+&{1\over 2}g_{AB}\biggl[
\left< B_{C}\right>\left< B_{D}\right>R^{(ind)}_{CD}
+2K\sigma\biggr]
\label{eqn:Einsteinvev1}
\end{eqnarray}
from $G_{AB},$ where we used also the previous results, namely the
$G_{NN}$ equation, the Israel matching condition and the $B_{A}$ equation.

Imposing that
$\nabla_{A}\left<B_{C}\right> =0,$ it follows that 
$\left<B_{AC}\right>=\nabla_{A}\left<B_{C}\right>
-\nabla_{C}\left<B_{A}\right> =0,$
which enables us to further simplify Eq.~(\ref{eqn:Einsteinvev1}):
\begin{eqnarray} 
&&{1\over \kappa_{(5)} ^2}\left[ 
G_{AB}^{(ind)} +2K_{AC}K_{BC} -{1\over 2}K_{AB}K
+{1\over 2}g_{AB}\left( R^{(ind)} -2K_{CD}K_{CD} -K^2\right)\right] 
+g_{AB}\Lambda_{(5)} \cr
&=&{1\over 2}\biggl[
{5\over 2}\left<B_{A}\right>\left<B_{C}\right>R^{(ind)}_{CB}
+{5\over 2}\left<B_{B}\right>\left<B_{C}\right>R^{(ind)}_{AC}
-6K_{AC}K_{BD}\left< B_{C}\right>\left< B_{D}\right>
+2K_{AB}~\sigma \biggr]\cr
&+&{1\over 2}g_{AB}\biggl[
\left< B_{C}\right>\left< B_{D}\right>R^{(ind)}_{CD}
+2K\sigma\biggr]~.
\label{eqn:Einsteinvev2}
\end{eqnarray}
Hence, in order to obtain a vanishing cosmological constant
and ensure that Lorentz invariance holds on the brane, we must impose
respectively that
\begin{equation}
\Lambda_{(5)} = %\kappa_{(5)} ^2 
K\sigma
\label{eqn:vanishingcc}
\end{equation} 
and that
\begin{eqnarray}
&&2K_{AC}K_{BC} -{1\over 2}K_{AB}K
+{1\over 2}g_{AB}\left( R^{(ind)} -2K_{CD}K_{CD} -K^2\right)
\cr
&=&\kappa_{(5)} ^2\biggl[
{5\over 4}\left<B_{A}\right>\left<B_{C}\right>R^{(ind)}_{CB}
+{5\over 4}\left<B_{B}\right>\left<B_{C}\right>R^{(ind)}_{AC}
+{1\over 2}g_{AB}
\left< B_{C}\right>\left< B_{D}\right>R^{(ind)}_{CD}\cr
&&-3K_{AC}K_{BD}\left< B_{C}\right>\left< B_{D}\right>
%+{1\over 2}g_{AB}
%\left< B_{C}\right>\left< B_{D}\right>R^{(ind)}_{CD}
+K_{AB}~\sigma \biggr]
%&&{}+{1\over 2}g_{AB}
%\left< B_{C}\right>\left< B_{D}\right>R^{(ind)}_{CD}\biggr].
\label{eqn:Lorentzbrane}
\end{eqnarray}
We observe that there is close relation between the vanishing of the 
cosmological constant and the keeping of the Lorentz invariance on the brane. 
These conditions are enforced so that the higher dimensional 
signatures encapsulated in the induced geometry of the brane cancel
the Lorentz symmetry breaking inevitably induced on the brane, thus
reproducing the observed geometry. The first condition, 
Eqn.~(\ref{eqn:vanishingcc}), can be modified to account for any
non-vanishing value for the cosmological constant, as appears to be 
suggested by the recent Wilkinson Microwave Anisotropy Probe 
(WMAP) data, by defining the observed
cosmological constant $\Lambda$ such that 
$\Lambda_{(5)}= \Lambda +\tilde \Lambda_{(5)}.$ 
A much elaborate fine-tuning, however, is required 
for the Lorentz symmetry to be observed on the brane, as described by 
the condition in Eqn.~(\ref{eqn:Lorentzbrane}).
To our knowledge this is 
a new feature in braneworld models, as in most models Lorentz
invariance is a symmetry  
shared by both the bulk and the brane. We shall examine    
further implications of this mechanism elsewhere \cite{BeCarvalho06}. In this 
future study the inclusion of a scalar field will also be discussed.

\section{Discussion and Conclusions}

In this paper we analyse the spontaneous symmetry breaking of
Lorentz invariance in the bulk and its consequent effect on the brane. 
For this purpose, we considered a bulk vector field subject to a
potential which endows the field with a non-vanishing  
vacuum expectation value, thus allowing for the spontaneous breaking
of the Lorentz symmetry. This bulk vector field is   
directly coupled to the Ricci tensor, so that after the breaking of
Lorentz invariance the loss of  
this symmetry is transmitted to the gravitational sector of the model. 
For simplicity, we assumed that the vacuum expectation 
value of the component of the vector field normal to the brane vanishes.
The complex interplay between matching conditions and the Lorentz symmetry 
breaking terms was examined. We found that Lorentz invariance 
on the brane can be made exact via the dynamics of the graviton, vector field
and the geometry of the extrinsic curvature of the surface of the 
brane. As a consequence of the exact reproduction of Lorentz symmetry on
the brane, we found a condition for the matching of the observed
cosmological constant in four dimensions.  
This tuning does not follow from a dynamical mechanism but 
is imposed by phenomenological reasons only. 
From this point of view, both the value of the cosmological constant
and the Lorentz symmetry seem to be a consequence of a complex fine-tuning. 
We aim to further study the implications of our mechanism by considering also 
the inclusion of a scalar field in a forthcoming publication 
\cite{BeCarvalho06}.  
    
%\vfill
\vskip 0.5cm

%%%%%%%%%%%%%%%%%%%%%%%%%%%%%%%%%%%%%%%%%%%%%%%%%%%%%%%%%%%%%%%%%%%%%%%%%
\centerline{\bf {Acknowledgments}}

\vskip 0.2cm

\noindent 
C. C. thanks Funda\c c\~ao para a Ci\^encia e a 
Tecnologia (Portuguese Agency) for financial support under the
fellowship /BPD/18236/2004. C. C. thanks Martin Bucher, Georgios
Kofinas and Rodrigo Olea for useful discussions, and the National and
Kapodistrian University of Athens for its hospitality.

%\vfill

%%%%%%%%%%%%%%%%%%%%%%%%%%%%%%%%%%%%%%%%%%%%%%%%%%%%%%%%%%%%%%%%%%%%%
%\newpage

\end{document}